\documentclass[twocolumn]{autart}    

\usepackage{graphicx}          
\usepackage{graphics} 
\usepackage{epsfig} 
\usepackage{mathptmx} 
\usepackage{times} 
\usepackage{amsmath} 
\usepackage{amssymb}  
\usepackage{floatrow}
\usepackage{enumerate}
\usepackage{mathrsfs}
\usepackage{amsfonts}

\DeclareMathOperator{\Tr}{Tr}
\DeclareMathOperator{\Dg}{Diag}
\DeclareMathOperator{\Sp}{Span}
\DeclareMathOperator{\Spec}{Spec}
\begin{document}

\begin{frontmatter}

\title{An Exponential Quantum Projection Filter for Open Quantum Systems \thanksref{footnoteinfo}} 

\thanks[footnoteinfo]{This research is supported in part by a Hong Kong Research Grant council (RGC) grant (No. 15206915), the Air Force Office of Scientific Research (AFOSR) and the Office of Naval Research Global (ONRS) under agreement number FA2386-16-1-4065, and the Australian Research Council under grant number DP180101805. Corresponding author G.~Zhang. Tel. +852 2766 6936.
Fax +852 2764 4382.}

\author[PolyU]{Qing Gao}\ead{qing.gao.chance@gmail.com},
\author[PolyU]{Guofeng Zhang}\ead{Guofeng.Zhang@polyu.edu.hk},
\author[ANU]{Ian R. Petersen}\ead{i.r.petersen@gmail.com}    

\address[PolyU]{Department of Applied Mathematics, Hong Kong Polytechnic University, Hong Kong SAR, China.} 
\address[ANU]{Research School of Engineering, Australian National University, Canberra, ACT, 2601, Australia.}


\begin{abstract}                          
An approximate exponential quantum projection filtering scheme is developed for a class of open quantum systems described by Hudson-Parthasarathy quantum stochastic differential equations, aiming to reduce the computational burden associated with online calculation of the quantum filter. By using a differential geometric approach, the quantum trajectory is constrained in a finite-dimensional differentiable manifold consisting of an unnormalized exponential family of quantum density operators, and an exponential quantum projection filter is then formulated as a number of stochastic differential equations satisfied by the finite-dimensional coordinate system of this manifold. A convenient design of the differentiable manifold is also presented through reduction of the local approximation errors, which yields a simplification of the quantum projection filter equations. It is shown that the computational cost can be significantly reduced by using the quantum projection filter instead of the quantum filter. It is also shown that when the quantum projection filtering approach is applied to a class of open quantum systems that asymptotically converge to a pure state, the input-to-state stability of the corresponding exponential quantum projection filter can be established. Simulation results from an atomic ensemble system example are provided to illustrate the performance of the projection filtering scheme. It is expected that the proposed approach can be used in developing more efficient quantum control methods.\\
\end{abstract}

\end{frontmatter}

\section{Introduction}
The past decades have witnessed tremendous advances in quantum technologies which allow us to effectively probe and manipulate matter at the level of atoms (e.g., \cite{Akimoto2011}, \cite{Viola2003}, \cite{Hamerly2012}, \cite{Liu2016}). A basic requirement in realizing these technologies is to infer the unknown quantum system states from measurements. Nevertheless, two fundamental nonclassical features manifested by quantum systems are that i) any quantum measurement scheme can extract in principle only partial information from the observed quantum system; and ii) any quantum measurement inevitably changes the quantum system states in a probabilistic way (\cite{Breuer2002}, \cite{Gardiner2000}, \cite{Wiseman2009}). As a result, any measurement based quantum feedback control problem is essentially a problem of stochastic control theory with partial observations and can generally be converted into a control problem for a quantum filter with fully accessible states, as in classical stochastic control theory (\cite{Handel2005a}, \cite{Handel2005c}, \cite{Mirrahimi2007},  \cite{Ticozzi2008}, \cite{Ticozzi2009}). In this context, the quantum system and observations are modelled as a pair of quantum stochastic differential equations, while the quantum filter, also known as the quantum trajectory, is a dynamic equation driven by the classical output signal of a laboratory measuring device (\cite{Belavkin1992}, \cite{Bouten2007}, \cite{Gao2016}). A quantum filter recursively updates the information state of a quantum system undergoing continual measurements and provides real-time information that can be fed into the quantum controller. Therefore, real time solution of the quantum filter equations is essential in implementing a quantum feedback control setup, which, however, tends to be computationally expensive, especially when the quantum system has a high dimension (\cite{Song2016}). 

In order to make the implementation more efficient, several approaches have been proposed in the literature concerning the approximation or model reduction of quantum filter, to mention a few, see (\cite{Emzir2016}, \cite{Rouchon2015}). In \cite{Emzir2016}, an extended Kalman filtering approach was developed for a class of open quantum systems subject to continuous measurement, where time-varying linearization was applied to the system dynamics and a Kalman filter was designed for the linearized system. The proposed approach performs well for nearly linear quantum systems. A numerical approach to reducing the computational burden associated with calculating quantum trajectories was discussed in \cite{Rouchon2015} and was used to demonstrate a two-qubit feedback control scheme. It was shown in simulation studies that a high approximation accuracy can be achieved even when a small number of integration steps is involved. 

The main goal of this paper is to approximate the optimal quantum filter using a lower-dimensional quantum projection filter, motivated by the pioneering work on projection filtering for classical stochastic systems by Brigo, Hanzon and LeGland (\cite{Brigo1998}, \cite{Brigo1999}). The basic idea of projection filtering is to constrain the optimal filter to remain in a finite-dimensional submanifold embedded in the state space of the filter. Then the projection filter can be expressed as a set of dynamic equations satisfied by the local coordinates of this submanifold. The problem of quantum projection filtering has been addressed in \cite{Handel2005b} where the information state of a highly nonlinear quantum model of a strongly coupled two-level atom in an optical cavity was approximately determined by a tractable set of stochastic differential equations. However, the approach in \cite{Handel2005b} requires exact prior knowledge of an invariant set of the solutions to the quantum filter equations. In other words, a finite-dimensional family of densities is already known to be a good approximation of the information state. This restrictive assumption was removed in \cite{Nielsen2009} where an unsupervised learning identification algorithm was developed to determine the structure of the submanifold. However, the identification algorithm itself could be time consuming when a more general and complex open quantum system is considered instead of the simple two-level quantum system in \cite{Nielsen2009}. In this paper, we design an \emph{exponential quantum projection filter} for a general atom-laser interaction system subject to continuous homodyne detection, using a differential-geometric method in quantum information geometry theory. We propose a finite-dimensional differentiable submanifold consisting of an unnormalized exponential family of quantum density operators, on which a quantum Fisher metric structure is rigorously defined. Then through a projection operation, the solutions to the unnormalized quantum filter equations are maintained in this submanifold. In other words, the resulting quantum trajectory becomes a curve on the finite-dimensional manifold and the unnormalized quantum filter equation reduces to a set of recursive equations satisfied on the corresponding finite-dimensional coordinate system. We also present a convenient design of the differentiable manifold, by which the local approximation errors are significantly reduced and the quantum projection filter equations are simplified. In addition, it is shown that when the projection filtering strategy is applied to a class of open quantum systems that asymptotically converge to a pure state, the input-to-state stability of the corresponding exponential quantum projection filter can be established.

This paper is organized as follows. Section 2 introduces some preliminaries on the quantum system model, quantum filter and quantum information geometry. Section 3 presents the main contributions of this paper. Here, we first derive the exponential quantum projection filter equations and provide a convenient design of the differentiable manifold in Subsection 3.1. Then we apply the projection filtering strategy to an asymptotically stable open quantum system and analyze the behaviour of the corresponding exponential quantum projection filter in Subsection 3.2. Section 4 applies the proposed approach to an atomic ensemble interacting with an electromagnetic field and demonstrates the approximation performance through simulation studies. Section 5 concludes this paper.

\textbf{Notation.} i=$\sqrt{-1}$. Here we use the Roman type character i to distinguish the imaginary unit from the index $i$. $A^{\dagger}$ represents the complex conjugate transpose of matrix $A$. $s_1(A),...,s_n(A)$ are the singular values of matrix $A$ which are arranged in decreasing order, i.e., $s_1(A)\geq s_2(A)\geq...\geq s_n(A)\geq 0$. $\Tr(A)$ is the trace of matrix $A$. $[A, B]=AB-BA$ is the commutator of matrices $A$ and $B$. $I$ is the identity matrix. $\|A\|$ is the max norm of matrix $A$. $\mathbb{R}^n$ and $\mathbb{C}^n$ represent the $n$-dimensional real vector space and complex vector space, respectively.

\section{Preliminaries}
\subsection{System Formulation and Quantum Filter}
We sketch the open quantum system model under consideration in this section; a more detailed description can be found in (\cite{Bouten2007}, \cite{Breuer2002}, \cite{Song2016}) and the references therein. 

In this paper, we consider a typical physical scenario from quantum optics. An arbitrary quantum system $G$, e.g., an atomic ensemble, is in weak interaction with an external single-channel laser field that is initially in the vacuum state. A cavity is used to increase the interaction strength between the light and the quantum system. One of the cavity mirrors, through which a forward mode of the electromagnetic field scatters off, is made slightly leaky such that information about the quantum system $G$ is extracted using a homodyne detector. The single-channel probe laser field has an annihilation operator $b(t)$ and a creation operator $b^{\dagger}(t)$, which are operators defined on a symmetric Fock space $\mathcal{E}$ that can be decomposed into the past and future components in the form of a tensor product $\mathcal{E}=\mathcal{E}_{t]}\otimes \mathcal{E}_{(t}$. Let $B(t)=\int_0^tb(s)ds$ and $B^{\dagger}(t)=\int_0^tb^{\dagger}(s)ds$ be integrated annihilation and creation field operators on $\mathcal{E}$, respectively. In this paper, the laser field is supposed to be canonical, that is, 
\begin{align*}
&dB(t)dB^{\dagger}(t)=dt, \\
&dB^{\dagger}(t)dB(t)=dB^{\dagger}(t)dB^{\dagger}(t)=dB(t)dB(t)=0.
\end{align*}
Let us denote by $\mathcal{H}_{\mathcal{Q}}$ the Hilbert space of the quantum system $G$ and suppose $\dim(\mathcal{H}_{\mathcal{Q}})=n<\infty$. The composite system composed of the atomic system and the field is assumed to be isolated. Then its temporal Heisenberg-picture evolution can be described by a unitary operator $U(t)$ on the tensor product Hilbert space $\mathcal{H}_{\mathcal{Q}}\otimes \mathcal{E}$, which satisfies the following Hudson-Parthasarathy quantum stochastic differential equation\footnote[1]{We have assumed $\hbar$=1 by using atomic units in this paper.}:
\begin{eqnarray*}
dU(t)=\left\{\left(-\mbox{i}H-\frac{1}{2}L^{\dagger}L\right)dt+LdB^{\dagger}(t)-L^{\dagger}dB(t)\right\}U(t)
\end{eqnarray*}
with the initial condition $U(0)=I$, where $H$ is the initial Hamiltonian of the quantum system $G$, and $L$ is a coupling operator, or measurement operator that describes how the system interact with the input field. The joint system state $\pi_0\otimes \left|\upsilon \right>\left<\upsilon \right|$ is given by some quantum state $\pi_0$ in $\mathcal{H}_{\mathcal{Q}}$ and the vacuum state $\left|\upsilon \right>$. 

In the Heisenberg picture, an initial system operator $X$ evolves to $j_t(X)=U^{\dagger}(t)(X\otimes I)U(t)$ at time $t$. Using the quantum $It\hat o$ rules, $j_t(X)$ satisfies the following quantum master equation:
\begin{align}
dj_t(X)=&j_t(\mathscr{L}_{L, H}(X))dt \nonumber\\
&+j_t([L^{\dagger},X])dB(t)+j_t([X,L])dB^{\dagger}(t), \label{gprojection1}
\end{align}
where $\mathscr{L}_{L, H}$ is the so-called Lindblad generator:
\begin{align}
\mathscr{L}_{L, H}(X)=\mbox{i}[H,X]+L^{\dagger}XL-\frac{1}{2}(L^{\dagger}LX+XL^{\dagger}L). \label{gprojection13}
\end{align}
A homodyne detector measures the observable $Y(t)=U^{\dagger}(t)Q(t)U(t)$ where $Q(t)=B(t)+B^{\dagger}(t)$ is the real quadrature of the input laser field and generates a classical photocurrent signal. The so-called \emph{self-nondemolition} property, i.e., $[Y(s), Y(t)]=0$ for all $s\leq t$ enables monitoring $Y(t)$ continuously and interpreting $Y(t)$ as a classical signal (photocurrent). By the $It\hat o$ rules, $Y(t)$ satisfies 
\begin{align}
dY(t)=U^{\dagger}(t)(L+L^{\dagger})U(t)dt+dQ(t).\label{gprojection2}
\end{align}
Equations (\ref{gprojection1}) and (\ref{gprojection2}) form the system-observation pair of our model. As in classical stochastic control theory, the goal of quantum filtering is to find the least-mean-square estimate of the system observable $j_t(X)$ given the prior observations $Y(s), 0\leq s \leq t$, that is, to derive an expression for the quantum conditional expectation $\pi_t(X)=\mathbb{E}(j_t(X)|\mathscr{Y}(t))$ where $\mathscr{Y}(t)$ is the commutative von Neumann algebra generated by the observation process $Y(s), 0\leq s \leq t$. From the so-called \emph{nondemolition} condition, i.e.,  $[j_t(X), Y(s)]=0$ for all $s\leq t$, $\pi_t(X)$ can be isomorphically interpreted as a classical conditional expectation and is thus well defined (\cite{Bouten2007}, \cite{Gao2016}). The dynamic equation satisfied by $\pi_t(X)$ has been derived as  (\cite{Belavkin1992}, \cite{Bouten2007}):
\begin{align}
d\pi_t(X)=&\pi_t(\mathscr{L}_{L, H}(X))dt+\left(\pi_t(L^{\dagger}X+XL)\right.\nonumber\\
&\left.-\pi_t(L^{\dagger}+L)\pi_t(X)\right)\left(dY(t)-\pi_t(L^{\dagger}+L)dt\right). \label{gprojection3}
\end{align}
In this paper, we are mainly concerned with the adjoint form of the quantum filter in (\ref{gprojection3}). Defining the conditional quantum density matrix $\rho_t$ by $\pi_t(X)=\Tr(\rho_tX)$, the filter equation in (\ref{gprojection3}) yields
\begin{eqnarray}
d\rho_t=\mathscr{L}_{L, H}^{\dagger}(\rho_t)dt+\mathscr{D}_{L}(\rho_t)\left(dY(t)-\Tr(\rho_t(L+L^{\dagger})dt\right), \label{gquantum1}
\end{eqnarray}
with $\rho_0=\pi_0$. Here $\mathscr{L}_{L, H}^{\dagger}$ is the adjoint Lindblad generator:
\begin{equation}
\mathscr{L}_{L, H}^{\dagger}(X)=-\mbox{i}[H,X]+LXL^{\dagger}-\frac{1}{2}(L^{\dagger}LX+XL^{\dagger}L), \label{gprojection12}
\end{equation}
and $\mathscr{D}_{L}(X)=LX+XL^{\dagger}-X\Tr(X(L+L^{\dagger})).$

Note that the quantum filter (\ref{gquantum1}) is a classical stochastic differential equation that is driven by the Wiener type classical photocurrent signal $Y(t)$ and can thus be conveniently implemented on a classical signal processor. Equation (\ref{gquantum1}) has been widely used in applications including quantum state estimation and quantum feedback control (\cite{Hamerly2012}, \cite{Handel2005a}, \cite{Handel2005c}), where in time calculation of (\ref{gquantum1}) is essential. However, one has to solve a system of $n^2-1$ recursive $It\hat o$ stochastic differential equations in order to determine the conditional probability density $\rho_t$ defined on $\mathcal{H}_{\mathcal{Q}}$. A high computational cost will arise if the atomic system has a large number of energy levels. The main goal of this paper is to reduce the dimension of the filtering equations while guaranteeing acceptable approximation performance.

\subsection{Quantum Information Geometry}
This subsection will introduce some foundations of quantum information geometry theory. A more detailed formulation can be found in Chapter 7 of the book (\cite{Amari2000}). Let the set of all self-adjoint operators on the Hilbert space $\mathcal{H}_{\mathcal{Q}}$ be denoted by
\begin{eqnarray}
\mathbb{A}=\{A|A=A^{\dagger}\}. \label{gquantum2}
\end{eqnarray}
Subsequently, we focus on the geometry of the set of nonnegative self-adjoint operators which is denoted by
\begin{eqnarray}
\mathbb{Q}=\{\rho|\rho\geq 0, \rho\in \mathbb{A}\}. \label{gquantum3}
\end{eqnarray}
Hence $\mathbb{Q}$ is a closed subset of $\mathbb{A}$ and is naturally regarded as a real manifold with dimension $\dim(\mathbb{Q})=n^2$. Apparently, the tangent space at each point $\rho$ to $\mathbb{Q}$, which is denoted by $\mathscr{T}_{\rho}(\mathbb{Q})$, is identified with $\mathbb{A}$. When a tangent vector $X\in \mathscr{T}_{\rho}(\mathbb{Q})$ is considered as an element of $\mathbb{A}$ by this identification, we denote it by $X^{(m)}$ and call it the \emph{mixture representation (m-representation)} of $X$. When a coordinate system $[\varepsilon^{i}], i=1,2,...,n^2,$ is given on $\mathbb{Q}$ so that each state is parameterised as $\rho_{\varepsilon}$, the $m-$representation of the natural basis vector of the tangent vector space is identified with
\begin{eqnarray}
(\partial_i)^{(m)}=\partial_i, \label{gquantum4}
\end{eqnarray}
where $\partial_i:=\partial \rho_{\varepsilon} /\partial \varepsilon^{i}$. Naturally, $\{\partial_i\}$ are linearly independent and
\begin{eqnarray}
\mathscr{T}_{\rho_{\varepsilon}}(\mathbb{Q})=\Sp\{\partial_i\}. \label{gquantum5}
\end{eqnarray}
A differentiable manifold is not naturally endowed with an inner product structure. We need to add a Riemannian structure to the manifold. To be specific, we define a Riemannian metric on $\mathbb{Q}$. The \emph{symmetrized inner product} is employed to define the inner product $\{\ll,\gg_{\rho}, \rho\in \mathbb{Q}\}$ on $\mathbb{A}$ (\cite{Amari2000}):
\begin{eqnarray}
\ll A,B\gg_{\rho}=\frac{1}{2}\Tr(\rho AB+\rho BA), \forall A, B\in \mathbb{A}. \label{gquantum6}
\end{eqnarray}
Based on this inner product, we define another useful representation called the $e-representation$ of a tangent vector $X\in \mathscr{T}_{\rho}(\mathbb{Q})$ as the self-adjoint operator $X^{(e)}\in \mathbb{A}$ satisfying
\begin{eqnarray}
\ll X^{(e)}, A\gg_{\rho}=\Tr\left(X^{(m)}A\right), \forall A\in \mathbb{A}. \label{gquantum7}
\end{eqnarray}
Using the $e-$representation defined above, we define an inner product $\left<,\right>$ on $\mathscr{T}_{\rho}(\mathbb{Q})$ by 
\begin{align}
\left<X,Y\right>_{\rho}&=\ll X^{(e)}, Y^{(e)}\gg_{\rho}\nonumber\\
&=\Tr\left(X^{(m)}Y^{(e)}\right),  \forall X, Y\in \mathscr{T}_{\rho}(\mathbb{Q}). \label{gquantum8}
\end{align}
Then $g=\left<,\right>$ forms a Riemmanian metric on $\mathbb{Q}$ which may be regarded as a quantum version of the Fisher metric. The components of this metric are given by
\begin{eqnarray}
g_{ij}=\left<\partial_i,\partial_j\right>_{\rho}=\Tr\left(\partial_i^{(m)}\partial_j^{(e)}\right). \label{gquantum9}
\end{eqnarray}
\textbf{Example 2.1.}  Consider a qubit system, the Hilbert space $\mathcal{H}_{\mathcal{Q}}$ is identified with $\mathbb{C}^2$. Denote
\begin{eqnarray}
Q_1=\frac{I+\sigma_z}{2},Q_2=\frac{I-\sigma_z}{2},Q_3=\sigma_x, \mbox{ and }Q_4=\sigma_y, \label{revision6}
\end{eqnarray}
where  $\sigma_x,\sigma_y,\sigma_z$ are Pauli matrices described by 
\begin{eqnarray}
\sigma_x=
\left(
\begin{array}{cc}
  0  & 1   \\
 1   & 0  
\end{array}
\right), \sigma_y=
\left(
\begin{array}{cc}
  0  & -i   \\
 i   & 0  
\end{array}
\right) \mbox{ and }\sigma_z=
\left(
\begin{array}{cc}
  1  & 0   \\
 0   & -1  
\end{array}
\right). \label{revision7}
\end{eqnarray}
Then $Q_i\in \mathbb{Q}\subset \mathbb{A}$. Each $\rho_{\varepsilon}\in \mathbb{Q}$ can be represented by
\begin{eqnarray}
\rho_{\varepsilon}=\sum_{i=1}^4\varepsilon^i Q_i. \label{revision8}
\end{eqnarray}
In this case, one has
\begin{eqnarray}
(\partial_i)^{(m)}=\partial_i=Q_i,\label{revision9}
\end{eqnarray}
and
\begin{eqnarray}
\mathscr{T}_{\rho_{\varepsilon}}(\mathbb{Q})=\Sp\{\partial_i\},\label{revision10}
\end{eqnarray}
respectively.
Then, given any $X\in \mathscr{T}_{\rho_{\varepsilon}}(\mathbb{Q})$, its $m-$representation $X^{(m)}$ is a linear combination of $Q_i$
and its $e-$representation can be derived from (\ref{gquantum7}).

\section{An Exponential Quantum Projection Filter: Design and Analysis}
In this section, we propose a projection filtering approach to approximating the quantum filter equation in (\ref{gquantum1}), using differential geometric methods in quantum information geometry theory. The basic idea of the projection filtering strategy is illustrated in Fig. 1. We consider to apply a projection operation to a space of unnormalized quantum density operators and map the optimal quantum filter equation onto a fixed lower-dimensional submanifold. A natural basis will be derived for the tangent space at each point of this submanifold, and a local projection operation can be defined with respect to a quantum Fisher metric to map the infinitesimal increments generated by the quantum filter equation onto such tangent spaces. The resulting stochastic vector field on the submanifold then defines the dynamics of the approximation filter. In this paper, we consider to use a submanifold consisting of an unnormalized exponential family of quantum density operators. It is noted that quantum density operators in the exponential form is useful in practice, examples being Gaussian states and general thermal states (\cite{Amari2000}, \cite{Jiang2014}).
\begin{figure}
\includegraphics[width=1\linewidth]{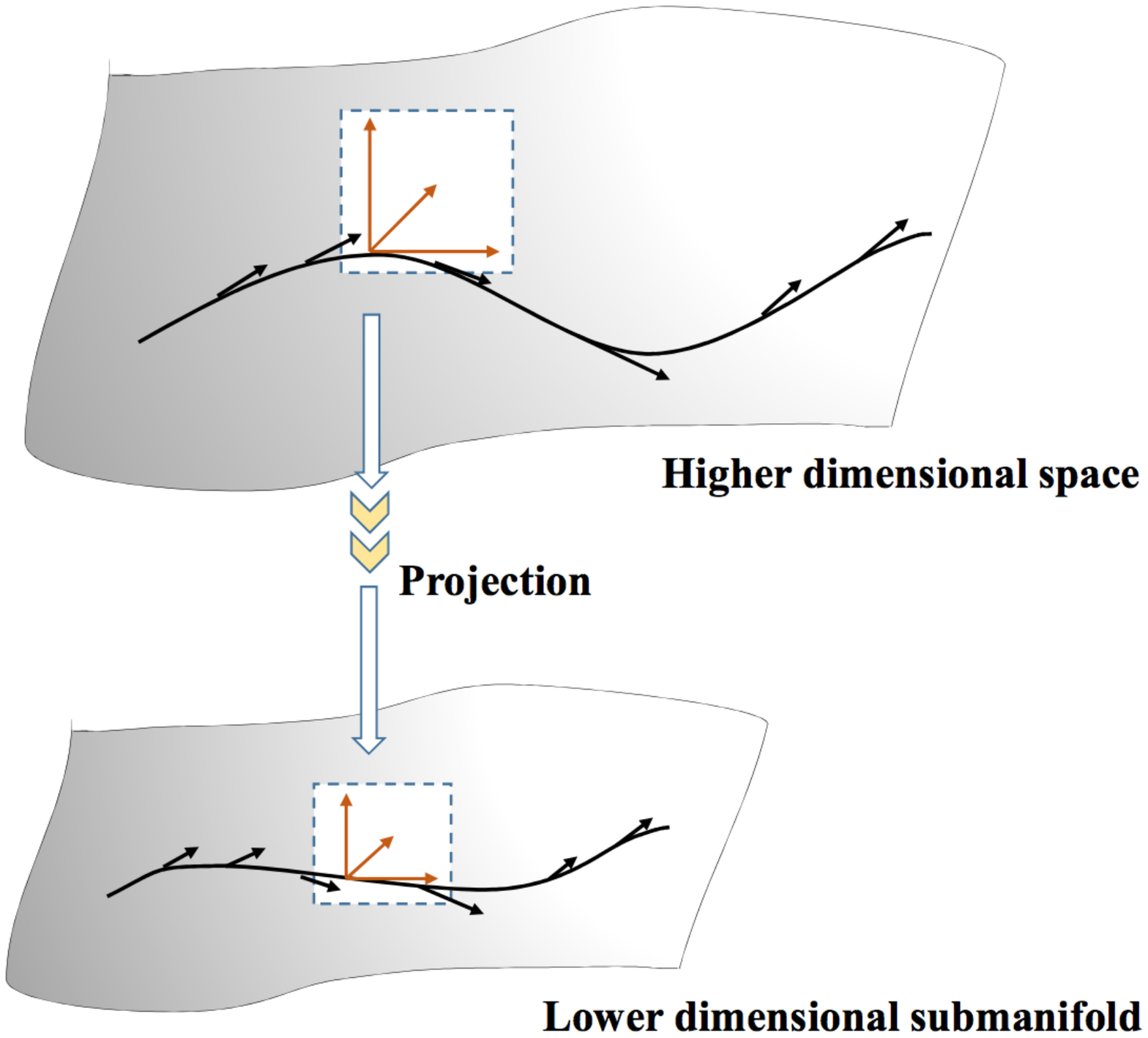}
\caption{Cartoon illustrating the basic ideas of projection filtering (\cite{Brigo1998}, \cite{Brigo1999}). The solid black line in the upper half of the figure represents a quantum trajectory which flows within a higher dimensional space. The vector field (the black arrows) along the curve represents the temporal filter dynamics. The tangent vectors at each point live in a linear space called the tangent vector space. In the bottom half of the figure, we build a lower dimensional submanifold which is embedded in the higher dimensional space. The goal is to make the solutions to the optimal filter remain in the finite dimensional submanifold. This can be done if we project the linear tangent vector space of the higher dimensional space to the tangent vector space of the lower dimensional one, and make the two solid lines start from the same initial point.}
\end{figure}

\subsection{Design of the Quantum Projection Filter}
The quantum projection filter equation will be derived in this subsection. We start from the unnormalized version of the quantum filter equation in (\ref{gquantum1}):
\begin{eqnarray}
d\bar \rho_t=\mathscr{L}_{L, H}^{\dagger}(\bar \rho_t)dt+\left(L\bar \rho_t+\bar \rho_tL^{\dagger}\right)dY(t), \label{gquantum10}
\end{eqnarray}
where $\bar \rho_t$ is the unnormalized information state corresponding to $\rho_t$ such that $\rho_t=\bar \rho_t/\Tr(\bar \rho_t)$. $\bar \rho_t$ is initially set to be $\bar \rho_0=\rho_0=\pi_0$. The unnormalized filter equation (\ref{gquantum10}) is used since its linear form is easier to manipulate compared with the nonlinear filter equation in (\ref{gquantum1}). 

It is worth mentioning that in order to illustrate the unnormalized quantum filter using a differential manifold structure, one must interpret the stochastic differential equation in (\ref{gquantum10})  using Stratonovich integral theory because $It\hat o$'s rule is incompatible with a manifold structure \cite{Brigo1998}. We have the following result.

\textbf{Lemma 3.1.} The $It\hat o$ quantum stochastic differential equation in (\ref{gquantum10}) can be equivalently rewritten as the following Stratonovich quantum stochastic differential equation:
\begin{eqnarray}
d\bar \rho_t=\left(-\mbox{i}[H, \bar \rho_t]-\mathscr{S}_{L}(\bar \rho_t)\right)dt+\left(L\bar \rho_t+\bar \rho_tL^{\dagger}\right)\circ dY(t), \label{gquantum11}
\end{eqnarray}
where
\begin{eqnarray}
\mathscr{S}_{L}(\bar \rho_t)=\frac{\left(L+L^{\dagger})L\bar \rho_t+\bar \rho_tL^{\dagger}(L+L^{\dagger}\right)}{2}. \label{gprojection4}
\end{eqnarray}
\emph{Proof.} The proof of Lemma 3.1 is given in Appendix.

Now we design the quantum projection filter following the scheme illustrated in Fig. 1. On one hand, it follows from (\ref{gquantum10}) that $\bar \rho_t$ is nonnegative and self-adjoint. Thus the totality of the unnormalized quantum density matrix $\bar \rho_t$ is identified with the set $\mathbb{Q}$ in (\ref{gquantum3}). It can be verified that the two terms $-i[H, \bar \rho_t]-\mathscr{S}_{L}(\bar \rho_t)$ and $L\bar \rho_t+\bar \rho_tL^{\dagger}$ on the right hand side of (\ref{gquantum11}) are vectors in $\mathscr{T}_{\rho}(\mathbb{Q})$, or equivalently, operators belonging to the set $\mathbb{A}$ in (\ref{gquantum2}).  On the other hand, the submanifold is designed to be a $C^{\infty}$ manifold consisting of an exponential family of unnormalized quantum density operators:
\begin{eqnarray}
\mathbb{S}=\{\bar \rho_{\theta}\}=\left\{e^{\frac{1}{2}\sum_{i=1}^m\theta_iA_i}\rho_0e^{\frac{1}{2}\sum_{i=1}^m\theta_iA_i}\right\}, \label{gquantum12}
\end{eqnarray}
where the submanifold operators $A_i\in \mathbb{A}, i\in\{1,2,...,m\}$ are mutually commutating and pre-designed.  We suppose that the entire submanifold $\mathbb{S}$ can be covered by a single coordinate chart $(\mathbb{S},\theta=(\theta_1,...,\theta_m)\in \Theta)$, where $\Theta$ is an open subset of $\mathbb{R}^{m}$ containing the origin. Then we have $\dim\{\mathbb{S}\}=m$. 

According to the chain rule in Stratonovich stochastic calculus, we have 
\begin{eqnarray}
d\bar \rho_{\theta}=\sum_{i=1}^m \bar \partial_i\circ d\theta_i,\label{gquantum13}
\end{eqnarray}
where $\bar \partial_i:=\partial \bar \rho_{\theta}/\partial \theta_i$. Assuming the set $\{\bar \partial_1,...,\bar \partial_m\}$ is linearly independent, then this set forms an $m-$representation of the natural basis of $\mathscr{T}_{\bar \rho_{\theta}}(\mathbb{S})$; i.e., the tangent vector space at each point $\bar \rho_{\theta}$ to $\mathbb{S}$. We have
\begin{eqnarray}
\mathscr{T}_{\bar \rho_{\theta}}(\mathbb{S})=\Sp\{\bar \partial_i, i=1,...,m.\}.\label{gquantum14}
\end{eqnarray}
A direct calculation using Stratonovich stochastic calculus yields
\begin{eqnarray}
\frac{\partial \bar \rho_{\theta}}{\partial \theta_i}=\frac{1}{2}(A_i\bar \rho_{\theta}+\bar \rho_{\theta}A_i). \label{gquantum15}
\end{eqnarray}
It then follows directly from (\ref{gquantum7}) and (\ref{gquantum15}) that $\bar \partial_i^{(e)}=A_i$. Thus each component of the quantum Fisher metric in (\ref{gquantum9}) is given by a real-valued function of $\theta$:
\begin{align}
g_{ij}(\theta)&=\ll \bar \partial_i^{(e)}, \bar \partial_j^{(e)} \gg_{\bar \rho_{\theta}}=\Tr(\bar \rho_{\theta}A_iA_j)\nonumber\\
&=\Tr\left(\rho_0e^{\frac{1}{2}\sum_{i=1}^m\theta_iA_i}A_iA_je^{\frac{1}{2}\sum_{i=1}^m\theta_iA_i}\right),\label{gquantum16}
\end{align}
because the operator $e^{\frac{1}{2}\sum_{i=1}^m\theta_iA_i}A_iA_je^{\frac{1}{2}\sum_{i=1}^m\theta_iA_i}$ is self-adjoint. The quantum Fisher information matrix is an $m\times m$ dimensional real matrix given by $G(\theta)=(g_{ij}(\theta))$. Then an orthogonal projection operation $\Pi_{\theta}$ can be defined for every $\theta\in \Theta$ as follows:
\begin{eqnarray}
\mathbb{A} &&\longrightarrow \mathscr{T}_{\bar \rho_{\theta}}(\mathbb{S}) \nonumber\\
\nu && \longmapsto \sum_{i=1}^m\sum_{j=1}^m g^{ij}(\theta) \left<\nu,\bar \partial_j\right>_{\bar \rho_{\theta}}\bar \partial_i, \label{gquantum17}
\end{eqnarray}
where the matrix $\left(g^{ij}(\theta)\right)$ is the inverse of the quantum information matrix $G(\theta)$.

Consider a curve in $\mathbb{S}$ around the point $\bar \rho_{\theta}$ to be of the form $\zeta: t \mapsto \bar \rho_{\theta_t}$. This corresponds to a real curve $\gamma: t \mapsto \theta_t$ in $\Theta$ around the real vector $\theta$, through the coordinate chart $(\mathbb{S},\theta)$. Let us consider that the curve $\zeta$ starts from the initial condition that  $\bar \rho_{\theta_0}=\pi_0$, or equivalently, the curve $\zeta$ starts from $\theta_0=0$. The unnormalized exponential quantum projection filter is then defined as the following quantum stochastic differential equation on the $m$-dimensional differentiable manifold $\mathbb{S}$:
\begin{align}
d\bar \rho_{\theta_t}&=\Pi_{\theta_t}\left(-\mbox{i}[H,\bar \rho_{\theta_t}]\right)dt+\Pi_{\theta_t}\left(-\mathscr{S}_{L}(\bar \rho_{\theta_t})\right)dt\nonumber\\
&\hspace{2cm}+\Pi_{\theta_t}\left(L\bar \rho_{\theta_t}+\bar \rho_{\theta_t}L^{\dagger}\right)\circ dY(t).\label{gquantum18}
\end{align}
From the definition of the manifold $\mathbb{S}$ in (\ref{gquantum12}), the projection quantum filter can be equivalently written using the equations satisfied by the real curve $\gamma$ in $\Theta$. Denote $\theta_t=(\theta_1(t),...,\theta_m(t))'$. An explicit form of the curve equations is given in the following theorem.

\textbf{Theorem 3.1.} The real curve $\gamma: t \mapsto \theta_t$ satisfies the following recursive stochastic differential equation:
\begin{eqnarray}
d\theta_t=G(\theta_t)^{-1}\{\Xi(\theta_t)dt+\Gamma(\theta_t)\circ dY(t)\}, \label{gquantum19}
\end{eqnarray}
with the initial conditions $\theta_i(0)=0, i=1,...,m$, where $\Xi(\theta_t)$ and $\Gamma(\theta_t)$ are both $m-$dimensional column vectors of real functions on $\theta_t$. The $j$th elements of these quantities are given by
\begin{eqnarray*}
\Xi_j(\theta_t)=\Tr\left(\bar \rho_{\theta_t}\left(\mbox{i}[H, A_j]-\frac{A_j(L+L^{\dagger})L+L^{\dagger}(L+L^{\dagger})A_j}{2}\right)\right),
\end{eqnarray*}
and 
\begin{eqnarray*}
\Gamma_j(\theta_t)=\Tr(\bar \rho_{\theta_t}(A_jL+L^{\dagger}A_j)),
\end{eqnarray*}
respectively. 

\emph{Proof.} Applying the projection operation in (\ref{gquantum17}) and the chain rule in (\ref{gquantum13}) to the filter equation (\ref{gquantum11}) yields
\begin{align*}
&d\bar \rho_{\theta_t}=\sum_{i=1}^m \bar \partial_i\circ d\theta_i(t) \nonumber\\
&=\sum_{i=1}^m\sum_{j=1}^m g^{ij}(\theta) \Tr((-\mbox{i}[H,\bar \rho_{\theta_t}]-\mathscr{S}_{L}(\bar \rho_{\theta_t}))A_j)\bar \partial_idt\nonumber\\
&\hspace{4mm}+\sum_{i=1}^m\sum_{j=1}^m g^{ij}(\theta) \Tr((L\bar \rho_{\theta_t}+\bar \rho_{\theta_t}L^{\dagger})A_j)\bar \partial_i\circ dY(t)
\end{align*}
\begin{align}
&=\sum_{i=1}^m\sum_{j=1}^m g^{ij}(\theta)\Tr(\bar \rho_{\theta_t}(\mbox{i}[H,A_j]-\mathscr{S}_{L}^{\dagger}(A_j)))dt\bar \partial_i\nonumber\\
&\hspace{4mm}+\sum_{i=1}^m\sum_{j=1}^m g^{ij}(\theta)\Tr(\bar \rho_{\theta_t}(A_jL+L^{\dagger}A_j))\circ dY(t)\bar \partial_i,\label{gquantum20}
\end{align}
where $\mathscr{S}_{L}^{\dagger}(X)=\frac{X(L+L^{\dagger})L+L^{\dagger}(L+L^{\dagger})X}{2}$ is the adjoint of the operator $\mathscr{S}_{L}$ in (\ref{gprojection4}). The differential equation in (\ref{gquantum19}) can be obtained by comparing the coefficients of the natural basis $\{\bar \partial_i\}$ from both sides of (\ref{gquantum20}).  $\hspace{4cm}\Box$

The stochastic differential equation (\ref{gquantum19}) combined with the equation (\ref{gquantum12}) determines the unnormalized projection quantum density operator. In this paper, (\ref{gquantum18}) or (\ref{gquantum19}) is called the \emph{quantum projection filter}. The approximate quantum information state $\tilde \rho_t$ can be then simply obtained as
\begin{eqnarray}
\tilde \rho_t=\bar \rho_{\theta_t}/\Tr(\bar \rho_{\theta_t}). \label{add01}
\end{eqnarray}
It can be observed that only a system of $m$ stochastic differential equations is needed to be calculated in order to determine $\tilde \rho_t$. Recall that one has to calculate a collection of $n^2-1$ stochastic differential equations in determining the information state $\rho_t$ in the quantum filter equation (\ref{gquantum1}). Thus the computational cost would be reduced significantly if the number $m$ is chosen to be small.

The above design procedure requires the predesign of the submanifold operators $A_i, i=1,...,m$. The remainder of this subsection will be devoted to a convenient design method for these self-adjoint operators through reduction of the local approximation errors. In fact, the proposed approximation scheme in Theorem 3.1 is implemented through two steps. First, the right-hand side of (\ref{gquantum11}) is evaluated at the current projection filter quantum density operator $\bar \rho_{\theta(t)}$ on $\mathbb{S}$, instead of the true density operator $\bar \rho_t$. However, the right-hand side vectors $-i[H, \bar \rho_t]$, $-\mathscr{S}_{L}(\bar \rho_t)$ and $L\bar \rho_t+\bar \rho_tL^{\dagger}$ will generally make the solutions leave the manifold $\mathbb{S}$. Thus a second approximation is made by projecting these vector fields onto the linear tangent vector space $\mathscr{T}_{\bar \rho_{\theta}}(\mathbb{S})$. In the remainder of this subsection, we will present a design of the submanifold $\mathbb{S}$ by considering the local errors for the quantum projection filter occurring in the \emph{second} approximation step at time $t$.

Following similar ideas as in \cite{Brigo1998}, we define at each point $\bar \rho_{\theta_t}$ the prediction residual as 
\begin{eqnarray}
\mathfrak{P}(t)=-\mbox{i}[H, \bar \rho_{\theta_t}]-\Pi_{\theta_t}(-\mbox{i}[H, \bar \rho_{\theta_t}]), \label{gquantum21}
\end{eqnarray}
and two correction residuals as
\begin{eqnarray}
\mathfrak{C}_1(t)=-\mathscr{S}_{L}(\bar \rho_{\theta_t})-\Pi_{\theta_t}(-\mathscr{S}_{L}(\bar \rho_{\theta_t})) \label{gquantum22}
\end{eqnarray}
and
\begin{eqnarray}
\mathfrak{C}_2(t)=L\bar \rho_{\theta_t}+\bar \rho_{\theta_t}L^{\dagger}-\Pi_{\theta_t}(L\bar \rho_{\theta_t}+\bar \rho_{\theta_t}L^{\dagger}),\label{gquantum23}
\end{eqnarray}
respectively.

Although it is not required in Theorem 3.1, the following assumption will be essential in the subsequent analysis in this paper. 

\textbf{Assumption 3.1.} The coupling operator is self-adjoint, i.e., $L=L^{\dagger}$.

This assumption is practically reasonable in many experimental settings; e.g., trapping a cold atomic ensemble in an optical cavity (\cite{Handel2005c}, \cite{Thomsen2002}). Since $L$ is self-adjoint, it admits a spectral decomposition $L=\sum_{i=1}^{n_0}\lambda_iP_{L_i}$, where $n_0\leq n$ is the number of nonzero eigenvalues of $L$, the set $\{\lambda_i\}$ contains all of the nonzero real eigenvalues of $L$, and $\{P_{L_i}\}$ is a set of projection operators that satisfies $P_{L_j}P_{L_k}=\delta_{jk}P_{L_k}$. Then one has the following result:

\textbf{Theorem 3.2.} The correction residuals $\mathfrak{C}_1$ and $\mathfrak{C}_2$ are both identically zero for all $t\geq 0$, if the submanifold in (\ref{gquantum12}) is designed according to 
\begin{align}
\begin{cases}
m=n_0,\\
A_i=P_{L_i}.
\end{cases}\label{gprojection11}
\end{align}
Moreover, the exponential quantum projection filter (\ref{gquantum19}) becomes 
\begin{eqnarray}
d\theta_t=G(\theta_t)^{-1}\Tr(\mbox{i}\bar \rho_{\theta_t}[H, A_j])dt-2\alpha dt+2\beta dY(t),\label{gquantum29}
\end{eqnarray}
where $\alpha=(\lambda_1^2,...,\lambda_{m}^2)'$ and $\beta=(\lambda_1,...,\lambda_{m})'$.

\emph{Proof.} From the definitions of the natural basis in (\ref{gquantum15}), the projection operation in (\ref{gquantum17}) and the correction residuals in (\ref{gquantum22}) and (\ref{gquantum23}), one has
\begin{align}
&\mathfrak{C}_1(t)=\Pi_{\theta_t}\left(L^2\bar \rho_{\theta_t}+\bar \rho_{\theta_t}L^2\right)-\left(L^2\bar \rho_{\theta_t}+\bar \rho_{\theta_t}L^2\right) \nonumber\\
&=\sum_{k=1}^m\lambda_k^2\left\{\Pi_{\theta_t}\left(P_{L_k}^2\bar \rho_{\theta_t}+\bar \rho_{\theta_t}P_{L_k}^2\right)-\left(P_{L_k}^2\bar \rho_{\theta_t}+\bar \rho_{\theta_t}P_{L_k}^2\right)\right\}\nonumber\\
&=\sum_{k=1}^m\lambda_k^2\left\{\Pi_{\theta_t}\left(A_k\bar \rho_{\theta_t}+\bar \rho_{\theta_t}A_k\right)-\left(A_k\bar \rho_{\theta_t}+\bar \rho_{\theta_t}A_k\right)\right\}\nonumber\\
&=\sum_{k=1}^m2\lambda_k^2\left\{\Pi_{\theta_t}\left(\bar \partial_k\right)-\bar \partial_k\right\}=0, \label{gprojection5}
\end{align}
and
\begin{align}
\mathfrak{C}_2(t)&=\Pi_{\theta_t}\left(L\bar \rho_{\theta_t}+\bar \rho_{\theta_t}L\right)-\left(L\bar \rho_{\theta_t}+\bar \rho_{\theta_t}L\right) \nonumber\\
&=\sum_{k=1}^m2\lambda_k\left\{\Pi_{\theta_t}\left(\bar \partial_k\right)-\bar \partial_k\right\}=0. \label{gprojection6}
\end{align}
Through the design method in Theorem 3.1, the components of the quantum Fisher metric in (\ref{gquantum16}) are given by
\begin{align}
g_{ij}(\theta)=\Tr(\bar \rho_{\theta}A_iA_j)=\delta_{ij}\Tr(\bar \rho_{\theta}A_i), i,j\in\{1,...,m\},\label{gprojection7}
\end{align}
and the quantum Fisher matrix $G(\theta)=(g_{ij}(\theta))$ becomes a diagonal matrix
\begin{align}
G(\theta)=\Dg\{\Tr(\bar \rho_{\theta}A_1),...,\Tr(\bar \rho_{\theta}A_m)\}.\label{gprojection8}
\end{align}
The $j$th elements of the vector functions $\Xi(\theta_t)$ and $\Gamma(\theta_t)$ in (\ref{gquantum19}) are given by
\begin{align}
\Xi_j(\theta_t)&=\Tr\left\{\bar \rho_{\theta_t}\left(\mbox{i}[H, A_j]-\left(A_jL^2+L^2A_j\right)\right)\right\}, \nonumber\\
&=\Tr\left(\mbox{i}\bar \rho_{\theta_t}[H, A_j]\right)-2\lambda_j^2\Tr(\bar \rho_{\theta}A_j),\label{gprojection9}
\end{align}
and
\begin{align}
\Gamma_j(\theta_t)=\Tr(\bar \rho_{\theta_t}(A_jL+LA_j))=2\lambda_j\Tr(\bar \rho_{\theta}A_j), \label{gprojection10}
\end{align}
respectively. Then (\ref{gquantum29}) can be concluded by substituting (\ref{gprojection8}), (\ref{gprojection9}) and (\ref{gprojection10}) into the filter equation (\ref{gquantum19}). $\hspace{1.5cm}\Box$

It has been shown in Theorem 3.1 that, by using the design scheme as in (\ref{gprojection11}), the correction residuals $\mathfrak{C}_1$ and $\mathfrak{C}_2$ are both eliminated while the prediction residual $\mathfrak{P}(t)$ still exists. In general, it is difficult to analyze $\mathfrak{P}(t)$ which depends on the trajectory of the quantum projection filter. However, in a special case, an upper bound of $\mathfrak{P}(t)$ can be derived and the exponential quantum projection filter (\ref{gquantum29}) can be further simplified. 

The unnormalized quantum filter (\ref{gquantum10}) and the exponential quantum filter (\ref{gquantum18}) are both driven by the classical photocurrent $Y(t)$ which is a Wiener process with bounded drift under some classical probability measure $\mathrm{P}$. Using Girsanov's theorem, however, one can always find a measure $\mathrm{P}'$ that is equivalent to $\mathrm{P}$ such that $Y(t)$ is a Wiener process with zero drift on the interval $[0,T]$, where $T>0$ is a fixed time called the final time (Page 458, \cite{Mirrahimi2007}). Let $\hat{\mathbb{E}}$ denote the expectation operation with respect to the measure $\mathrm{P}'$. One has the following result.

\textbf{Theorem 3.3.} When $[H, L]=0$, if the submanifold in (\ref{gquantum12}) is designed according to (\ref{gprojection11}), the exponential quantum projection filter (\ref{gquantum29}) becomes 
\begin{eqnarray}
d\theta_t=-2\alpha dt+2\beta dY(t),\label{revision1}
\end{eqnarray}
and the correction residuals $\mathfrak{C}_1$ and $\mathfrak{C}_2$ are both identically zero for all $t\geq 0$. Moreover, the prediction residual $\mathfrak{P}(t)$ satisfies
\begin{eqnarray}
\hat{\mathbb{E}}\sqrt{\Tr(\mathfrak{P}(t)^2)}\leq \sqrt{\Tr(X_0^2)}, t\geq 0,\label{revision2}
\end{eqnarray}
where $X_0=-\mbox{i}[H, \rho_0]$.

\emph{Proof.} Since $[H, L]=0$ and $A_i=P_{L_i}$ is the projection operator of $L$, one has $[H, A_i]=0, i=1,2,...m$. Then the evolution of the coordinate system $\theta_t$ in (\ref{gquantum29}) reduces to a set of independent $It\hat o$ stochastic differential equations in (\ref{revision1}).

Next we prove (\ref{revision2}). Denote $\Lambda(t)=\frac{1}{2}\sum_{i=1}^m\theta_i(t)A_i$. Then the submanifold (\ref{gquantum12}) can be rewritten as $\mathbb{S}=\{\bar \rho_{\theta}\}=\left\{e^{\Lambda(t)}\rho_0e^{\Lambda(t)}\right\}$ and
\begin{eqnarray}
&&\mathfrak{P}(t)=-\mbox{i}[H, \bar \rho_{\theta_t}]-\Pi_{\theta_t}(-\mbox{i}[H, \bar \rho_{\theta_t}])\nonumber\\
&=&e^{\Lambda(t)}X_0e^{\Lambda(t)}-\sum_{i=1}^m\sum_{j=1}^m g^{ij}(\theta) \Tr(e^{\Lambda(t)}X_0e^{\Lambda(t)}A_j)\bar \partial_i, \nonumber\\
&=&e^{\Lambda(t)}X_0e^{\Lambda(t)}+\mbox{i}\sum_{i=1}^m\sum_{j=1}^m g^{ij}(\theta) \Tr(\bar \rho_{\theta}[A_j, H])\bar \partial_i,\nonumber\\
&=&e^{\Lambda(t)}X_0e^{\Lambda(t)}.\label{revision3}
\end{eqnarray}
It then follows from Lemma A2 in the Appendix that 
\begin{eqnarray}
&&\hat{\mathbb{E}}\sqrt{\Tr(\mathfrak{P}(t)^2)}=\hat{\mathbb{E}}\sqrt{\Tr(e^{2\Lambda(t)}X_0e^{2\Lambda(t)}X_0)}\nonumber\\
&\leq&\hat{\mathbb{E}}\sqrt{\sum_{i=1}^ms_i(e^{2\Lambda(t)}X_0e^{2\Lambda(t)}X_0)}\nonumber\\
&\leq& \hat{\mathbb{E}}\sqrt{\sum_{i=1}^ms_i^2(e^{2\Lambda(t)})s_i^2(X_0)}\nonumber\\
&\leq&\sqrt{\sum_{i=1}^ms_i^2(X_0)}\hat{\mathbb{E}}s_1(e^{2\Lambda(t)})=\sqrt{\Tr(X_0^2)}\max \limits_i\hat{\mathbb{E}}e^{\theta_i(t)}.\label{revision4}
\end{eqnarray}
By using the $It\hat o$ rules, one can calculate from (\ref{revision1}) that
\begin{align}
de^{\theta_i(t)}&=-2\lambda_i^2e^{\theta_i(t)}dt+\frac{1}{2}e^{\theta_i(t)}(2\lambda_i)^2dt+2\lambda_ie^{\theta_i(t)}dY(t) \nonumber\\
&=2\lambda_ie^{\theta_i(t)}dY(t), \label{revision5}
\end{align}
which implies that $\hat{\mathbb{E}}\left(e^{\theta_i(t)}\right)\equiv \hat{\mathbb{E}}\left(e^{\theta_i(0)}\right)=1$. Then (\ref{revision2}) can be concluded from (\ref{revision4}) and (\ref{revision5}). The proof is thus completed. \hspace{6.5cm}$\qed$

Under some conditions, the exponential quantum projection filter could be an exact expression for the quantum filter (\ref{gquantum10}). The following model reduction result is a corollary of Theorem 3.2.

\textbf{Corollary 3.1.} When the system Hamiltonian $H=0$, $\bar \rho_{\theta_t}\equiv \bar \rho_t$ if the submanifold is designed according to (\ref{gprojection11}).

\subsection{Practical Stability of the Quantum Projection Filter}
In this subsection, we will analyze the time behaviour of the  exponential quantum projection filter on the interval $[0,T]$. Before proceeding, the following notation is introduced. 

Let $\mathscr{O}$ be an orthonormal basis of $\mathcal{H}_{\mathcal{Q}}$. For the quantum filter equation in (\ref{gquantum1}) and for any $\psi\in \mathscr{O}$, let
\begin{align}
T_{\psi}=\inf\{t\geq 0|\rho_t=\left|\psi\right>\left<\psi\right|\}. \label{gp1}
\end{align}
\textbf{Definition 3.1.} (\cite{Benoist2014}) The quantum filter (\ref{gquantum1}) fulfils a nondemolition condition if there exists an orthonormal basis $\mathscr{O}$ such that for any $\psi\in \mathscr{O}$
\begin{align}
\Tr(\rho_t\left|\psi\right>\left<\psi\right|)=1, \forall t\geq T_{\psi}. \label{gp2}
\end{align}
The stable states $\left|\psi\right>\left<\psi\right|$, $\psi\in \mathscr{O}$, are called pointer states of the quantum filter.

Let $\left|\psi_0\right>\left<\psi_0\right|$ be a particular pointer state of the quantum filter (\ref{gquantum1}). Then the Hilbert space $\mathcal{H}_{\mathcal{Q}}$ can be decomposed in the direct sum $\mathcal{H}_{\mathcal{Q}}=\mathcal{H}_S\oplus \mathcal{H}_R$ where $\mathcal{H}_S=\mathbb{C}\left|\psi_0\right>$. This yields a convenient decomposition of all matrices on $X\in \mathcal{H}_{\mathcal{Q}}$, that is, by choosing an appropriate basis, $X$ can be written as
\begin{align}
X=\left(\begin{matrix}
X_S & X_P  \\
X_Q & X_R 
\end{matrix}\right), \label{gp3}
\end{align}
where $X_S, X_R, X_P$ and $X_Q$ are operators from $\mathcal{H}_S$ to $\mathcal{H}_S$, $\mathcal{H}_R$ to $\mathcal{H}_R$, $\mathcal{H}_R$ to $\mathcal{H}_S$, and $\mathcal{H}_S$ to $\mathcal{H}_R$, respectively.

Denote $\bar P_S=\left(\begin{matrix}
I & 0  \\
0 & 0 
\end{matrix}\right)$ and $\bar P_R=\left(\begin{matrix}
0 & 0  \\
0 & I
\end{matrix}\right)$ the orthogonal projectors on $\mathcal{H}_S$ and $\mathcal{H}_R$, respectively.

\textbf{Definition 3.2.} The quantum filter (\ref{gquantum1}) is said to be strongly globally asymptotically stable ($SGAS$), if it fulfils a nondemolition condition for an orthonormal basis $\mathscr{O}$ and there is a pointer state $\left|\psi_0\right>\left<\psi_0\right|\in \mathscr{O}$ such that, $\forall \rho_0$,
\begin{align}
\lim \limits_{t \to \infty} \|\rho_t-\bar P_S \rho_t \bar P_S\|=0, \mbox{a.s.} \label{gp4}
\end{align}
Similar to (\ref{gprojection13}), define an operation from $\mathcal{H}_R$ to $\mathcal{H}_R$:
\begin{align}
&\mathscr{L}_{L_R, H_R}(X_R)\nonumber\\
&=\mbox{i}[H_R,X_R]+L_R^{\dagger}X_RL_R-\frac{1}{2}(L_R^{\dagger}L_RX_R+X_RL_R^{\dagger}L_R). \label{gp5}
\end{align}
and denote its spectral abscissa as:
\begin{equation}
\Delta_0:=\min\left\{-\mbox{Re}(\lambda)|\lambda \in \Spec\left(\mathscr{L}_{L_R, H_R}\right)\right\}. \label{gp6}
\end{equation}
The following lemma can be concluded directly from Theorem 3 in (\cite{Benoist2014}) and Lemma 2.7 in (\cite{Benoist2017}).

\textbf{Lemma 3.2.} The quantum filter (\ref{gquantum1}) is $SGAS$, if and only if $[H, L]=0$ and $\Delta_0>0$.

In addition, the following useful lemma is introduced.

\textbf{Lemma 3.3.} (\cite{Benoist2017}) For any constant scalar $\epsilon>0$, there exists an operator $K_R>0$ on $\mathcal{H}_R$ such that
\begin{align}
\mathscr{L}_{L_R, H_R}(K_R)\leq -(\Delta_0-\epsilon)K_R. \label{gp7}
\end{align}
We are ready to present the main result of this subsection.

\textbf{Theorem 3.4.} Suppose a continuously monitored open quantum system modelled by (\ref{gprojection1}) and (\ref{gprojection2}) has a quantum filter (\ref{gquantum1}) which is $SGAS$. Then for any positive scalar $\epsilon$ such that $\Delta_0>\epsilon>0$ there exists a positive operator $K_R>I$ such that the solution to the exponential quantum projection filter (\ref{gquantum18}) satisfies:
\begin{align}
\Tr(\bar P_R\bar \rho_{\theta_t})\leq& \left(\Tr\left(K_R\right)\Tr\left(\bar P_R\rho_0\right)-\frac{\Tr(K_R)}{\Delta_0-\epsilon}s_1(X_0)\right)e^{-(\Delta_0-\epsilon)t}\nonumber\\
&+\frac{\Tr(K_R)}{\Delta_0-\epsilon}s_1(X_0), \mbox{a.s.}, \label{gp8}
\end{align}
where $X_0=-\mbox{i}[H,\rho_0]$ as defined in Theorem 3.3.

\emph{Proof.} Since the quantum filter (\ref{gquantum1}) is $SGAS$, it is implied from Lemma 3.2 that $[H, L]=0$ and $\Delta_0>0$. By designing the submanifold according to (\ref{gprojection11}), it follows from Theorems 3.2 and 3.3 that the correction residuals defined in (\ref{gquantum22}) and (\ref{gquantum23}) vanish, and the prediction residual in (\ref{gquantum21}) becomes
\begin{align}
\mathfrak{P}(t)&=e^{\frac{1}{2}\sum_{i=1}^m\theta_i(t)A_i}X_0e^{\frac{1}{2}\sum_{i=1}^m\theta_i(t)A_i}. \label{gp9}
\end{align}
The unnormalized exponential quantum projection filter (\ref{gquantum18}) can be rewritten as 
\begin{align}
d\bar \rho_{\theta_t}&=\left\{-\mbox{i}[H,\bar \rho_{\theta_t}]-\mathscr{S}_{L}(\bar \rho_{\theta_t})\right\}dt-\mathfrak{P}(t)dt\nonumber\\
&\hspace{2cm}+\left(L\bar \rho_{\theta_t}+\bar \rho_{\theta_t}L\right)\circ dY(t),  \label{gp11}
\end{align}
which can be converted into an $It\hat o$ stochastic differential equation using Lemma 1:
\begin{align}
d\bar \rho_{\theta_t}&=\mathscr{L}_{L, H}^{\dagger}(\bar \rho_{\theta_t})dt-\mathfrak{P}(t)dt+\left(L\bar \rho_{\theta_t}+\bar \rho_{\theta_t}L\right)dY(t). \label{gp12}
\end{align}
Let $\hat \rho_{\theta_t}=\hat{\mathbb{E}}(\bar \rho_{\theta_t})$ and $\hat{\mathfrak{P}}(t)=\hat{\mathbb{E}}(\mathfrak{P}(t))$. Also, let
$\hat \rho_{\theta_t}=\left(\begin{matrix}
(\hat  \rho_{\theta_t})_S & (\hat  \rho_{\theta_t})_P  \\
(\hat  \rho_{\theta_t})_Q & (\hat  \rho_{\theta_t})_R
\end{matrix}\right), \rho_0=\left(\begin{matrix}
(\rho_0)_S & (\rho_0)_P  \\
(\rho_0)_Q & (\rho_0)_R 
\end{matrix}\right)$
and
$ \hat{\mathfrak{P}}(t)=\left(\begin{matrix}
\mathfrak{P}_S(t) & \hat{\mathfrak{P}}_P(t)  \\
\mathfrak{P}_Q(t) & \hat{\mathfrak{P}}_R(t)
\end{matrix}\right)$
be the decompositions of $\hat \rho_{\theta_t}$, $\rho_0$ and $\hat{\mathfrak{P}}(t)$ corresponding to the subspace decomposition $\mathcal{H}_{\mathcal{Q}}=\mathcal{H}_S\oplus \mathcal{H}_R$, respectively. A direct calculation on (\ref{gp12}) yields
\begin{align}
d(\hat  \rho_{\theta_t})_R=\left\{\mathscr{L}_{L_R, H_R}^{\dagger}((\hat  \rho_{\theta_t})_R)-\hat{\mathfrak{P}}_R(t)\right\}dt, \label{gp13}
\end{align}
which has a solution 
\begin{align}
(\hat  \rho_{\theta_t})_R=e^{t\mathscr{L}_{L_R, H_R}^{\dagger}}(\rho_0)_R-\int_0^te^{(t-s)\mathscr{L}_{L_R, H_R}^{\dagger}}\hat{\mathfrak{P}}_R(s)ds. \label{gp14}
\end{align}
Let $K=\left(\begin{matrix}
0 & 0  \\
0 & K_R 
\end{matrix}\right)$ and $V_K(\bar \rho_{\theta_t})=\Tr\left(K\hat \rho_{\theta_t}\right)$. Then it follows from (\ref{gp14}) that
\begin{align}
&V_K(\bar \rho_{\theta_t})=\Tr\left(K\hat \rho_{\theta_t}\right)=\Tr\left(K_R(\hat  \rho_{\theta_t})_R\right)\nonumber\\
&=\Tr\left(e^{t\mathscr{L}_{L_R, H_R}^{\dagger}}(\rho_0)_RK_R\right)-\int_0^t\Tr\left(e^{(t-s)\mathscr{L}_{L_R, H_R}^{\dagger}}\hat{\mathfrak{P}}_R(s)K_R\right)ds\nonumber\\
&=\Tr\left(e^{t\mathscr{L}_{L_R, H_R}}K_R(\rho_0)_R\right)-\int_0^t\Tr\left(e^{(t-s)\mathscr{L}_{L_R, H_R}}K_R\hat{\mathfrak{P}}_R(s)\right)ds. \label{gp15}
\end{align}
On one hand, because $e^{t\mathscr{L}_{L_R, H_R}}$ is a strictly positive map and $K_R>0$, it follows that $e^{t\mathscr{L}_{L_R, H_R}}K_R>0$ (\cite{Evans1978}). Then from Lemma 3.3 and Lemma A1 in Appendix, one has that by choosing $\epsilon<\Delta_0$,
\begin{align}
\Tr\left(e^{t\mathscr{L}_{L_R, H_R}}K_R(\rho_0)_R\right)&\leq\Tr\left(e^{t\mathscr{L}_{L_R, H_R}}K_R\right)\Tr\left((\rho_0)_R\right)\nonumber\\
&\leq e^{-(\Delta_0-\epsilon)t}\Tr\left(K_R\right)\Tr\left((\rho_0)_R\right).\label{gp16}
\end{align}
One the other hand, based on Theorem 3.3, Lemma 3.3 and Lemma A2 in Appendix,
\begin{align}
&-\Tr\left(e^{(t-s)\mathscr{L}_{L_R, H_R}}K_R\hat{\mathfrak{P}}_R(s)\right)\nonumber\\
&\leq \sum_{i=1}^{n-1}s_i\left(e^{(t-s)\mathscr{L}_{L_R, H_R}}K_R\hat{\mathfrak{P}}_R(s)\right)\nonumber\\
&\leq \sum_{i=1}^{n-1}s_i\left(e^{(t-s)\mathscr{L}_{L_R, H_R}}K_R\right)s_i\left(\hat{\mathfrak{P}}_R(s)\right)\nonumber\\
&\leq \Tr\left(e^{(t-s)\mathscr{L}_{L_R, H_R}}K_R\right)s_1\left(\hat{\mathfrak{P}}_R(s)\right)\nonumber\\
&\leq e^{-(\Delta_0-\epsilon)(t-s)}\Tr\left(K_R\right)\hat{\mathbb{E}}\left\{s_1^2\left(\sum_{i=1}^m e^{\frac{1}{2}\theta_i(t)}A_i\right)\right\}s_1(X_0)\nonumber\\
&=e^{-(\Delta_0-\epsilon)(t-s)}\Tr\left(K_R\right)s_1(X_0)\max \limits_i\hat{\mathbb{E}}\left(e^{\theta_i(t)}\right) \nonumber\\
&=e^{-(\Delta_0-\epsilon)(t-s)}\Tr\left(K_R\right)s_1(X_0). \label{gp17}
\end{align}
Then one has that
\begin{align}
&-\int_0^t\Tr\left(e^{(t-s)\mathscr{L}_{L_R, H_R}}K_R\hat{\mathfrak{P}}_R(s)\right)ds\nonumber\\
&\leq \int_0^te^{-(\Delta_0-\epsilon)(t-s)}ds\Tr\left(K_R\right)s_1(X_0)\nonumber\\
&=s_1(X_0)\frac{\Tr\left(K_R\right)}{\Delta_0-\epsilon}\left(1-e^{-(\Delta_0-\epsilon)t}\right). \label{gp19}
\end{align}
In addition, it is noted that the inequality (\ref{gp7}) still holds when multiplying $K_R$ by any positive scalar. Thus one can choose $K_R \geq I$ such that 
\begin{align}
V_K(\bar \rho_{\theta_t})=\Tr\left(K\hat \rho_{\theta_t}\right)\geq \Tr(\bar P_R\hat \rho_{\theta_t})=\hat{\mathbb{E}}\left|\Tr(\bar P_R\bar \rho_{\theta_t})\right|,\label{gp20}
\end{align}
because $\Tr(\bar P_R\bar \rho_{\theta_t})$ is strictly positive.  Then (\ref{gp8}) can be concluded from (\ref{gp16}), (\ref{gp19}) and (\ref{gp20}). $\hspace{3.5cm}\Box$

It is noted that $\Tr(\bar P_R\bar \rho_{\theta_t})$ in (\ref{gp8}) serves as a linear Lyapunov function candidate for the subspace $\mathcal{H}_S=\mathbb{C}\left|\psi_0\right>$ (See Theorem 1.2 in (\cite{Benoist2017})). Thus Theorem 3.3 can be treated as an input-to-state stability result for the exponential quantum projection filter. When $[H, \rho_0]=0$, or equivalently, $X_0\equiv 0$, Theorem 3.3 reduces to an asymptotical stability result. 

As in Corollary 3.1, we have the following alternative model reduction result:

\textbf{Corollary 3.2.} Suppose $[H, \rho_0]=0$ and the quantum filter (\ref{gquantum1}) fulfils a nondemolition condition, then $\bar \rho_{\theta_t}\equiv \bar \rho_t$ if the submanifold is designed according to (\ref{gprojection11}).

\section{Example: A spin-$J$ system with dispersive coupling}
The illustrative physical model used consists of an ensemble of $N$ atomic spins interacting dispersively with an electromagnetic laser field (\cite{Mirrahimi2007}, \cite{Thomsen2002}, \cite{Handel2005a}). Here $N$ is a positive integer. The atomic sample is placed in a strongly driven and heavily damped optical cavity and all atomic transitions probing state  $\left|1\right>$ are assumed to be detuned far from the cavity resonance. A homodyne detector is used to continuously monitor the light scattered from the cavity. Similar setups have been exploited in experiments producing spin squeezed states which have practical applications in several metrology tasks like magnetometers (\cite{Stockton2004}) and atomic clocks (\cite{Wineland1994}). 

The collective properties of $N$ two-level atoms can be conveniently described by a spin$-J$ system, i.e., a collection of $N=2J$ spin-$\frac{1}{2}$ systems (\cite{Thomsen2002}). Let the internal states, $\left|0\right>$ and $\left|1\right>$, of each atom be the degenerate two-level ground state. The collective spin operators are given by $J_{\alpha}=\sum_{k=1}^N\frac{1}{2}\sigma_{\alpha}^{(k)} (\alpha=x,y,z)$, where $\sigma_{\alpha}^{(k)}=I^{\otimes (k-1)}\otimes \sigma_{\alpha} \otimes I^{\otimes (N-k)}$ are the Pauli operators for each particle.

Consider applying a magnetic field with regulatable strength $u(t)$ in the $y$-direction. $u(t)$ can either act as a control signal that depends on the output of the quantum filter in a measurement based feedback control scheme, or an external disturbance signal in quantum estimation. Assuming that the cavity has a sufficiently large decay rate, the cavity dynamics can be adiabatically eliminated (\cite{Thomsen2002}, \cite{Handel2005a}, \cite{Wiseman1993}) and the temporal evolution of this spin-$J$ system can be described by the following quantum stochastic differential equation:
\begin{align*}
dU(t)=&\left\{\left(-\mbox{i}u(t)J_y-\frac{1}{2}\mu J_z^2\right)dt\right.\\
&\hspace{2cm}\left.+\sqrt{\mu}J_z(dB^{\dagger}(t)-dB(t))\right\}U(t),
\end{align*}
where $\mu$ is the effective coupling strength. The dispersive interaction between the atomic ensemble and the laser field introduces a phase shift of the cavity field that linearly depends on the total population difference, a quantity characterized by the self-adjoint operator $J_z$ since $N$ is conserved. A continuous and noisy observation of $J_z$ can be accomplished through a quantum nondemolition (QND) measurement of the field observable $B(t)+B^{\dagger}(t)$, that is 
\begin{align*}
dY(t)=2\sqrt{\mu}U^{\dagger}(t)J_zU(t)dt+dQ(t),
\end{align*}
as in (\ref{gprojection2}). The corresponding quantum filter conditioned on the observation process $Y(s), 0\leq s \leq t$ is given by
\begin{align}
d\rho_t=&\mathscr{L}_{\sqrt{\mu}J_z, u(t)J_y}^{\dagger}(\rho_t)dt \nonumber\\
&\hspace{1cm}+\mathscr{D}_{\sqrt{\mu}J_z}(\rho_t)\left(dY(t)-2\sqrt{\mu}\Tr(\rho_tJ_z)\right). \label{gp21}
\end{align}
Calculating the filter equation in (\ref{gp21}) is generally equivalent to solving a collection of $4^N-1$ stochastic differential equations, which is computational expensive even when the number of atoms in the ensemble is small. In this simulation, we consider the simplest case that $N=2$, that is, a two-qubits system with dispersive coupling (\cite{Mirrahimi2007}). In this case, the solution to a total of \textbf{15} stochastic differential equations is generally necessary. Nevertheless, based on Theorem 3.2, one can approximately calculate the information state $\rho_t$ using an exponential quantum projection filter consisting of only \textbf{2} stochastic differential equations. To be specific, the submanifold $\mathbb{S}$ in (\ref{gquantum12}) can be designed according to (\ref{gprojection11}). That is, the dimension of the manifold is chosen to be $m=2$ and the two submanifold operators are given by
\begin{align*}
A_1=P_{L_1}=\Dg(1,0,0,0), \mbox{ } A_2=P_{L_2}=\Dg(0,0,0,1)
\end{align*}
respectively, where $P_{L_1}$ and $P_{L_2}$ are the two projection operators of $\sqrt{\mu}J_z$ corresponding to its two nonzero eigenvalues though spectral decomposition. The exponential quantum projection filter is given by (\ref{gquantum29}) with $\lambda_1=\sqrt{\mu}/2$ and $\lambda_2=-\sqrt{\mu}/2$. For the system state, we assume that initially the first atom is prepared to be in $0.25*\left|1\right>\left<1\right|+0.75*\left|0\right>\left<0\right|$ and the second atom is prepared to be in $0.5*\left|1\right>\left<1\right|+0.5*\left|0\right>\left<0\right|$. 

In the simulation, the photocurrent is simulated from $dY(t)=2\sqrt{\mu}\Tr(\rho_tJ_z)dt+dW(t)$ and is used to drive the exponential quantum projection filter.  Monte Carlo simulations have been conducted by using the discretization approach as in \cite{Higham2001}. The simulation parameters used are as follows: the simulation interval is $t\in [0,T]$ with $T=1$, the normally distributed variance is $\delta t=T/N_0$ with $N_0=2^{12}$, and the step size is chosen to be $\Delta t=2\delta t$. The effective coupling strength $\mu$ is set to be $\mu=1$. The disturbance signal, as shown in Fig. 2, is set to be $u(t)=5e^{-5t}a$ where $a$ is a random variable with standard normal distribution.

\begin{figure}
\includegraphics[width=1\linewidth]{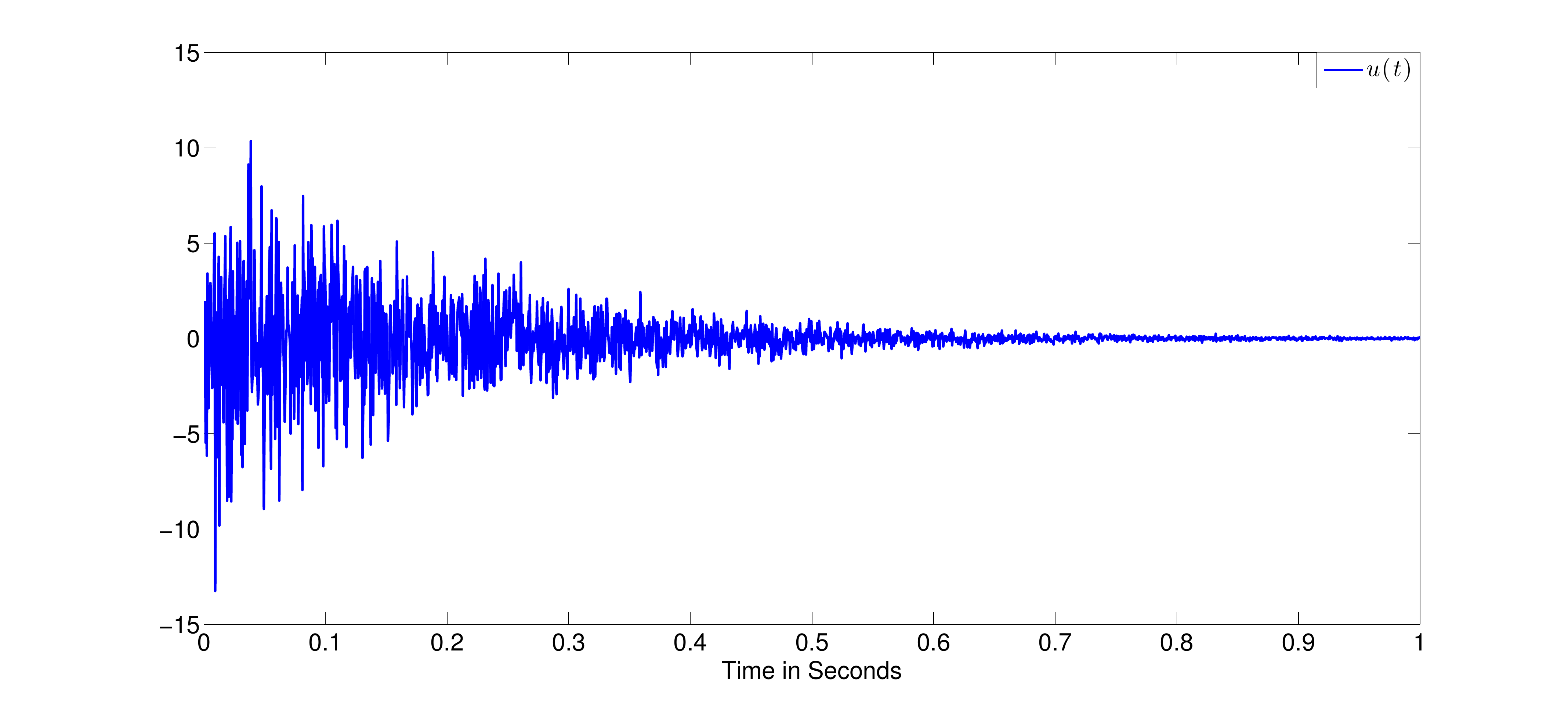}
\caption{The disturbance signal $u(t)$.}
\end{figure}

\begin{figure}
\includegraphics[width=1\linewidth]{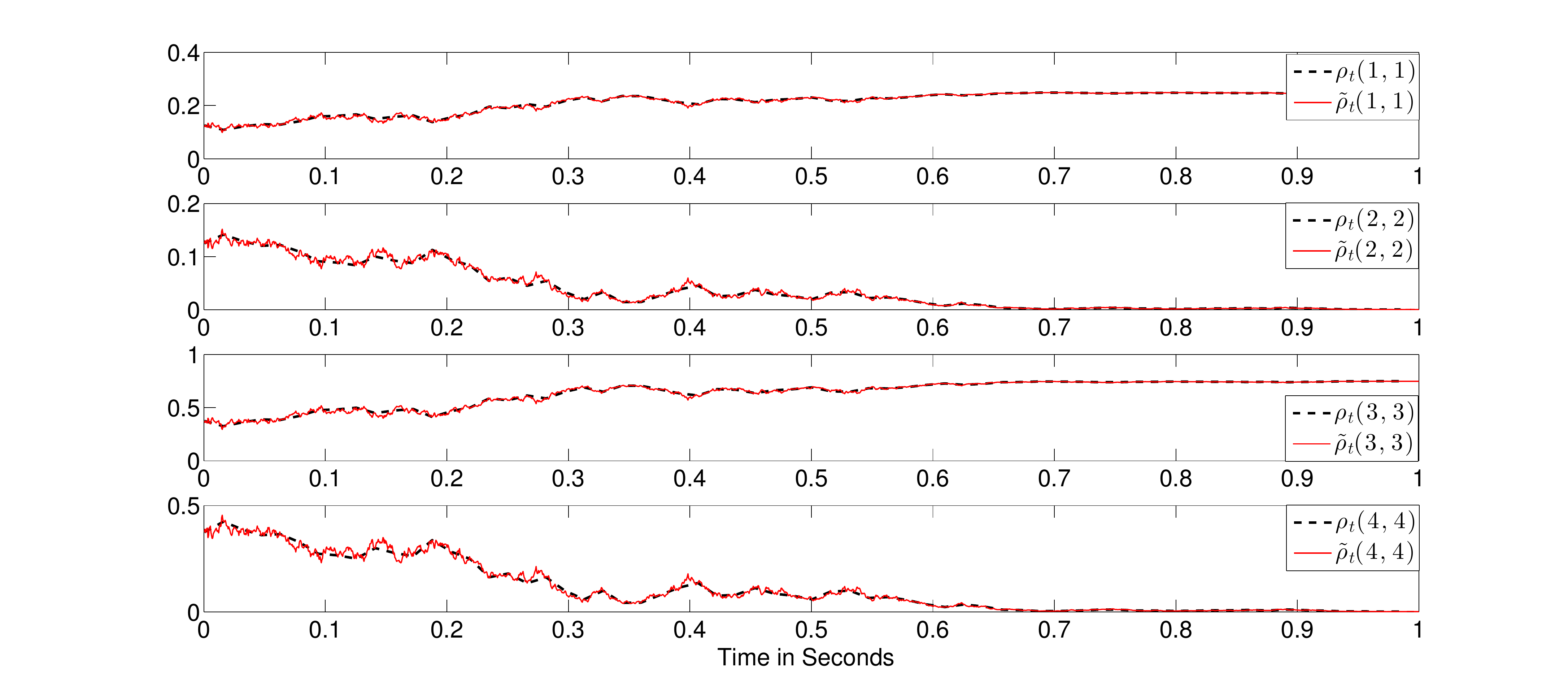}
\caption{The probabilities that each qubit is in its one particular internal states}
\end{figure}

\begin{figure}
\includegraphics[width=1\linewidth]{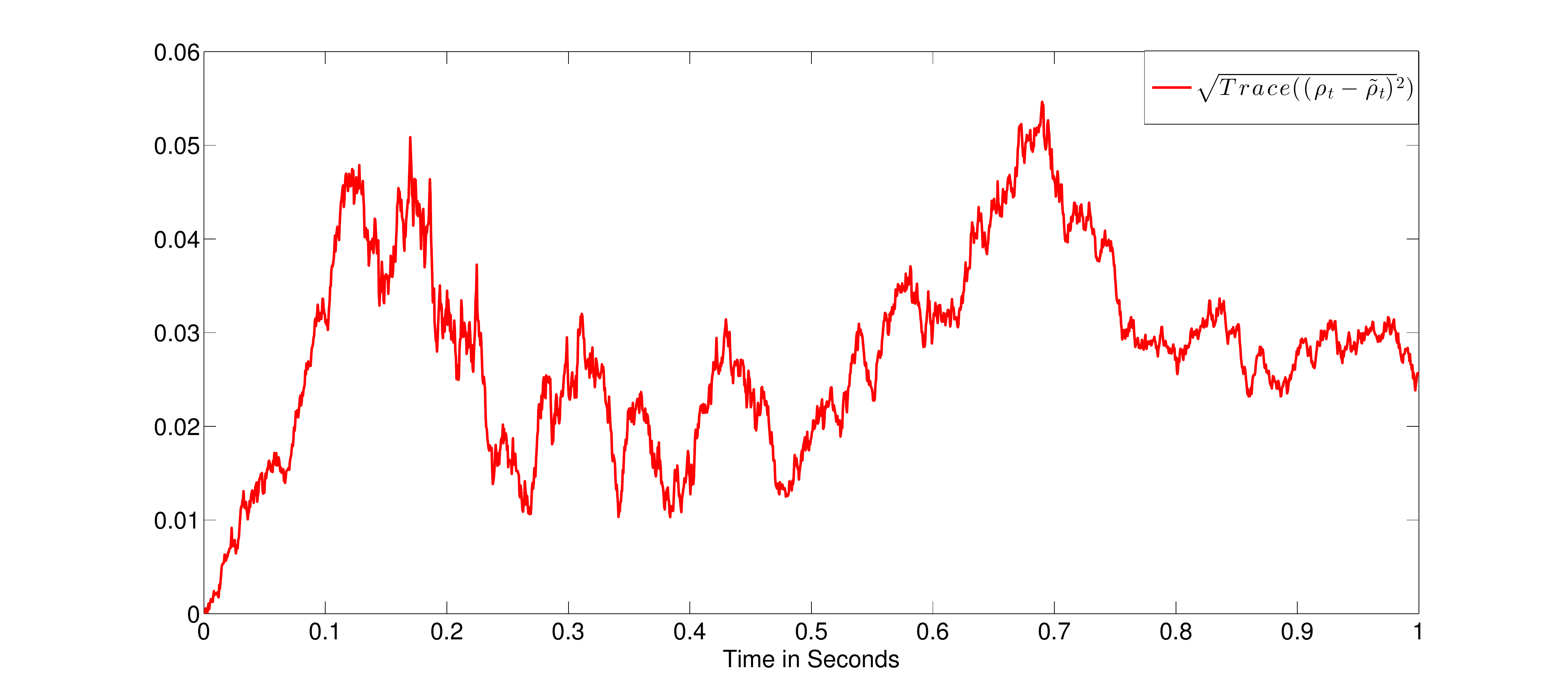}
\caption{Approximation error between the quantum filter and the quantum projection filter}
\end{figure}

The projection filtering strategy in Theorem 3.2 allows us to approximate the quantum information state $\rho_t$ by $\tilde \rho_t=\bar \rho_{\theta_t}/\Tr(\bar \rho_{\theta_t})$. The approximation performance of the proposed approximation filtering scheme is demonstrated by comparing the probabilities that each qubit is in its one particular internal states, calculated from the quantum filter equation in (\ref{gp21}) and the exponential quantum projection filter equation in (\ref{gquantum29}), respectively. In other words, we compare the trajectories of the process $\rho_t(1,1), \rho_t(2,2), \rho_t(3,3)$ and $\rho_t(4,4)$ with $\tilde \rho_t(1,1), \tilde \rho_t(2,2), \tilde \rho_t(3,3)$ and $\tilde\rho_t(4,4)$, respectively, which are depicted in the four subfigures of Fig. 3. The Frobenius norm of the difference between $\rho_t$ and $\tilde \rho_t$, i.e., $\sqrt{\Tr((\rho_t-\tilde \rho_t)^2)}$ is shown in Fig. 4. One can observe that $\rho_t$ and $\tilde \rho_t$ are very close over this time interval. This implies that in feedback control of the atomic ensemble one may use $\tilde \rho_t$ instead of $\rho_t$ in the controller in order to achieve a computationally more efficient design, although there might be small reduction of the control performance because of the approximation errors as shown in Fig. 3. Simulations also show that when the disturbance signal is set to be $u(t)\equiv 0$, then $\rho_t\equiv\tilde \rho_t$, which coincides with Corollary 3.1. 

In order to illustrate the stability analysis result in Theorem 3.4, we further consider the case that the external magnetic field with strength $u(t)$ is applied to the atomic ensemble in the $z$-direction. In this case, the system Hamiltonian is given by $H=u(t)J_z$ and thus commutes with the measurement operator $L=\sqrt{\mu}J_z$. All other parameters and settings remain the same and the simulation results are shown in Fig. 5. One can observe that $\rho_t$ in the filter (66) converges asymptotically to the target state $\rho_{tg}=\Dg(0,0.25,0,0.75)$, while the corresponding quantum projection filter is input-to-state stable, which is as expected from Theorem 3.4.

\begin{figure}
\includegraphics[width=1\linewidth]{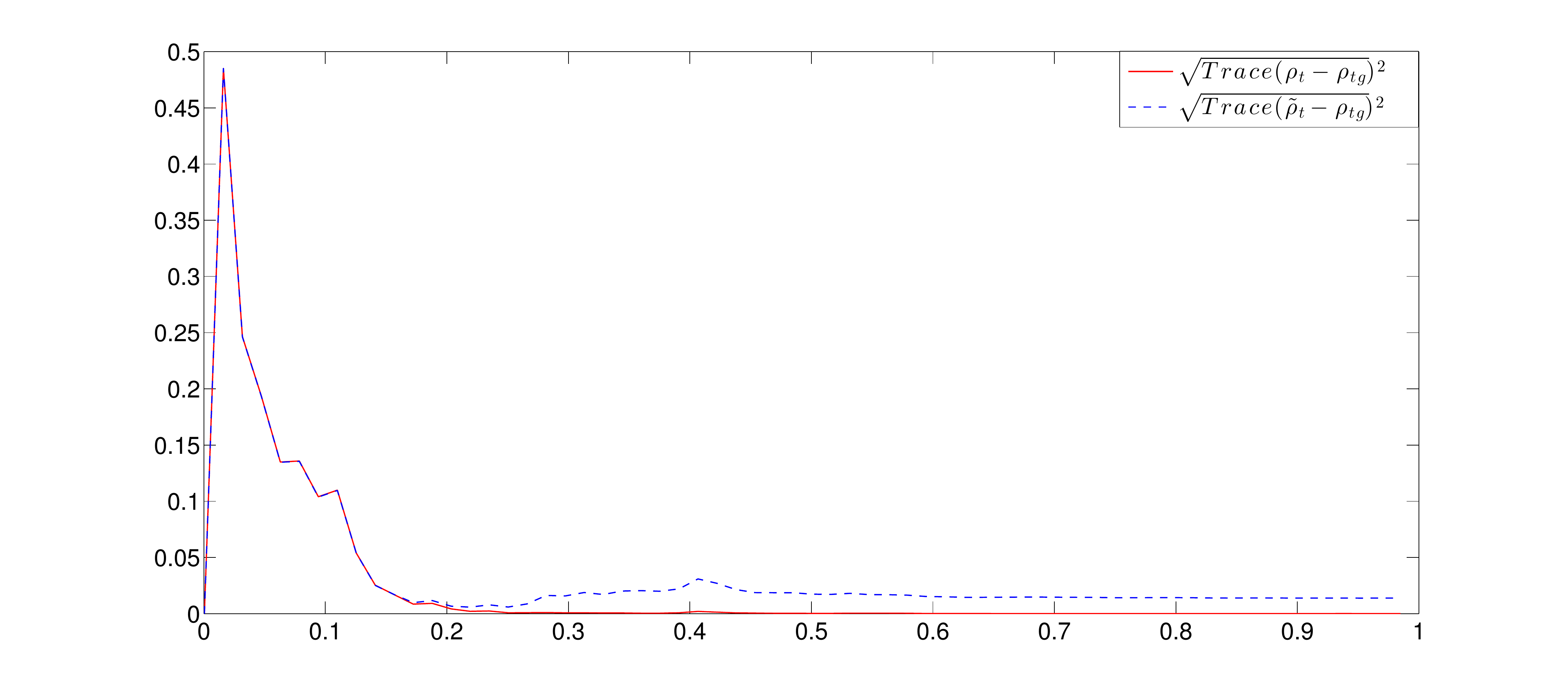}
\caption{Convergence characteristics of the quantum filter and the quantum projection filter, when applying a magnetic field with strength $u(t)$ in the $z$-direction}
\end{figure}

\section{Conclusions}
In this paper, a quantum projection filtering strategy is developed for a class of open quantum systems subject to homodyne detection. An exponential quantum projection filter is derived by defining a Riemann metric structure on a manifold consisting of an exponential family of unnormalized quantum density operators, which enables more efficient calculation of the quantum information state. The convergence capability of the exponential quantum projection filter for a special class of open quantum systems is also discussed. Simulations from an atomic ensemble with dispersive coupling show that the exponential quantum projection filter is able to approximate the quantum filter with a high accuracy. A number of open problems that deserve further research efforts are summarized as follows:
\begin{itemize}
\item Design of a submanifold such that the approximation errors are minimized;
\item Application of the quantum projection filter to feedback control of quantum systems;
\item Quantum projection filtering for open quantum systems with multiple Lindblad operators; and
\item Extension of the approach to the case of infinite dimensional quantum systems.
\end{itemize}

\section*{APPENDIX}

\textbf{Lemma A1}. (Page 269, \cite{Breitenbecker1972}). If $A$ and $B$ are positive semidefinite matrices, then
\begin{align}
0\leq \Tr(AB) \leq \Tr(A)\Tr(B).
\end{align}
\textbf{Lemma A2}. (Page 177, \cite{Cohen1988}). Let $A$ and $B$ are $n\times n$ matrices, Then,
\begin{align}
\sum_{i=1}^ks_i(AB)\leq \sum_{i=1}^ks_i(A)s_i(B), 1\leq k \leq n.
\end{align}
\emph{Proof of Lemma 3.1.}
Let $t_0<t_1<t_2...<t_p<T$ be a partition of any time interval $[t_0,T]$ and let the positive integer $p$ be big enough. A direct discretization of the filter equation (\ref{gquantum10}) yields
\begin{align}
\bar \rho_{t_{i+1}}&\simeq \bar \rho_{t_i}+\left(-i[H, \bar \rho_t]-\mathscr{S}_{L}(\bar \rho_t)\right)\Delta t_i\nonumber\\
&\hspace{4mm}+\left(L\bar \rho_t+\bar \rho_tL^{\dagger}\right)\Delta Y(t_i), i=0,1,...,p-1,\label{app1}
\end{align}
where $\Delta t_i=t_{i+1}-t_i$ and $\Delta Y(t_i)=Y(t_{i+1})-Y(t_i)$.

It is noted that $Y(t)$ is a classical Wiener process. Thus, when $p\to \infty$, one has $\Delta Y(t_i)\Delta Y(t_i)=\Delta t_i$ and $\Delta Y(t_i) \Delta t_i=0$. From the definition of the Stratonovich integral and (\ref{app1}), one has
\begin{align}
&(s) \int_{t_0}^T L\bar \rho_t+\bar \rho_t L^{\dagger}\circ dY(t)\nonumber\\
&=\lim\limits_{p\to \infty} \sum_{k=0}^p \frac{L(\bar \rho_{t_{k+1}}+\bar \rho_{t_{k}})+(\bar \rho_{t_{k+1}}+\bar \rho_{t_{k}})L^{\dagger}}{2}\Delta Y(t_{k})\nonumber\\
&=(I) \int_{t_0}^T L\bar \rho_t+\bar \rho_t L^{\dagger}dY(t)\nonumber\\
&\hspace{4mm}+\lim\limits_{p\to \infty} \sum_{k=0}^p \frac{L(L\bar \rho_{t_k}+\bar \rho_{t_{k}}L^{\dagger})+(L\bar \rho_{t_{k+1}}+\bar \rho_{t_{k}}L^{\dagger})L^{\dagger}}{2}\Delta t(t_{k})\nonumber\\
&=(I) \int_{t_0}^T L\bar \rho_t+\bar \rho_t L^{\dagger}dY(t)\nonumber\\
&\hspace{4mm}+\frac{1}{2}\int_{t_0}^TLL\bar \rho_t+\bar \rho_tL^{\dagger}L^{\dagger}+2L\bar \rho_tL^{\dagger}dt. \label{app2}
\end{align}
Lemma 3.1 can be obtained by substituting (\ref{app2}) into (\ref{gquantum10}). $\Box$

\section*{ACKNOWLEDGMENT}

Discussions with Prof. Haidong Yuan and Prof. Bo Qi are very much appreciated.

\end{document}